\documentclass[11pt, oneside]{article}   	% use "amsart" instead of "article" for AMSLaTeX format
\usepackage[T1]{fontenc}
\usepackage{lmodern}
\usepackage{geometry}                		% See geometry.pdf to learn the layout options. There are lots.
\usepackage{graphicx}				% Use pdf, png, jpg, or eps§ with pdflatex; use eps in DVI mode
\usepackage{amsmath}		
\usepackage{amssymb,amsfonts,amsthm}
\usepackage{mathtools,cool}
\usepackage{braket}

\title{\textbf{\Large Spectrum of Massive and Massless Ambitwistor Strings}}
\author{Christian Kunz \\ \small{\textit{E-mail:} \href{mailto:kunz.christian.321@gmail.com}{kunz.christian.321@gmail.com}}}
\usepackage{plain}
\usepackage{hyperref}
\hypersetup{
    colorlinks = true,
    linkcolor = {black},
    citecolor = {black},
    urlcolor={blue},
}

\newcommand{\ud}{\mathrm{d}}
\numberwithin{equation}{section}
\allowdisplaybreaks
\begin{document}
  \maketitle
  \tableofcontents

\begin{abstract}
  Inspired by recent work {\href{https://arxiv.org/abs/2301.11227}{arXiv:2301.11227}} on massive ambitwistor strings this paper examines the spectrum of such models using oscillator expansions. The spectrum depends heavily on the constant related to the normal ordering of the zero mode operator ${L_0}$ of the Virasoro algebra. The supergravity model is investigated in more detail, and two anomaly-free variations are presented, both with a rich spectrum and with tree scattering amplitudes that include a kinematic Parke-Taylor factor for particles other than gravitons without a need for an external current algebra. The spectrum of some of the models can also be interpreted as containing three generations of the Pati-Salam model.\\
\end{abstract}
\clearpage

\section{Introduction}
     The authors in \cite{Geyer:2018, Albonico:2020, Geyer:2020, Albonico:2022, Albonico:2023} examined dimensional reductions to 5 and 4 dimensions of a 6-dimensional massless ambitwistor string model. The resulting models in 4 dimensions contain two twistors describing massive particles. When equipped with maximal supersymmetry the models in 5 and 4 dimensions exhibit anomaly cancellation for its little group SL(2, $\mathbb{C}$) and also zero central charge for the Virasoro algebra if one assumes a central charge contribution of $c_{5d} = 10$ or $c_{6d} = 12$ arising from 5 or 6 compactified dimensions, respectively. Using the vertex operators provided in above references the genus zero worldsheet correlation functions lead to expected manifestly supersymmetric n-particle tree amplitudes.\\
     
     These ambitwistor string models use worldsheet spinors and, therefore, the strings are either in the Neveu-Schwarz (NS) or Ramond (R) sector, depending on their oscillator expansion. The actual spectrum then depends on the value $a$ of the zero mode $L_0$ of the Virasoro algebra when applied to a physical state. $a$ is determined by the Virasoro commutator $[L_n, L_{-n}]$ when all fields, including ghosts and antighosts, are taken into account:
     
     \begin{equation}
     [L_n, L_m] = (n - m) L_{n+m} + (\frac{c}{12} n^3 - 2 a n) \delta_{m+n,0}
     \label{virasoro}
     \end{equation}
     \\
     
     It will be shown that the vertex operators in \cite{Geyer:2020, Albonico:2022, Albonico:2023} implicitly assume that in \eqref{virasoro} $a = 0$ in the R sector and $a = 1$ in the NS sector. It follows that in the R sector only the zero modes of the twistor fields and in the NS sector only the $-\frac{1}{2}$ modes can contribute to physical states. For instance, this implicit assumption can be confirmed to happen when the extra dimensions are compactified with help of scalar fields (see appendix A of \cite{Albonico:2023}). \\
     
     On the other hand, one could also compactify extra components of the supertwistor and it will be shown that this can be done in such a way that in the massless limit it results in a supersymmetric ambitwistor string model with complete anomaly cancellation and a value of $a = 1$ in \eqref{virasoro} for both the R and NS sector. This model seems to have a richer spectrum than the previous model (in section 5 it will be shown that this is actually not really true), with non-zero twistor modes contributing to physical states in both the NS and R sector\footnote{The twistor modes always appear in pairs such that physical states and vertex operators on-shell are actually unaffected by the nature of background spin structures.}. The spectrum from modes in the NS sector leads to the same vertex operators as in the massless limit of the previous model, but the spectrum from modes in the R sector has some resemblance to the spectrum of the Berkovits-Witten twistor string with N=4 supersymmetry \cite{Berkovits_1:2004, Dolan:2008}, the main difference being that the new model has the additional SL(2,$\mathbb{C}$) little group symmetry, and it does not contain the 'dipole' states thought of being responsible for the lack of unitarity in the Berkovits-Witten model of conformal supergravity\cite{Berkovits_1:2004}. One important feature of the new model is that tree scattering amplitudes of spin $\le1$ particles contain a kinematic Parke-Taylor factor, without having to introduce an external current algebra. The model also has a non-supersymmetric variation with basically the same spectrum, except that the little group representation does not need to be the same for all the components of the multiplet. This non-supersymmetric version does not use auxiliary fields and was actually previously found independently by the author in \cite{Kunz:2020, Kunz_1:2020}.\\
     
     Unfortunately, because of the non-zero twistor modes in the R sector of the twistor components, these two models have complicated vertex operators, with unappealing scattering amplitudes. This motivated to find a model extension that keeps the Parke-Taylor factor for spin $\le1$ particle scattering and stays anomaly-free but now with a vanishing $L_0$ constant in the R sector. Then the spectrum nearly stays the same, as can be verified with a special interpretation of the spectrum of the $\mathcal{N} = 8$ supersymmetry. Actually, the same interpretation can be applied to the spectrum of the original model of \cite{Albonico:2022, Albonico:2023} showing that it already includes the same rich spectrum.\\
     
     The paper is organized as follows.\\
     In section \ref{Notation} the notation used in \cite{Albonico:2023} is reviewed and supplemented with the oscillator expansion for all the supertwistor component fields.\\
     In section \ref{Spectrum} it is shown that the vertex operators in \cite{Geyer:2020, Albonico:2022, Albonico:2023} indicate a spectrum that only involves the zero supertwistor modes in the R sector and the $-\frac{1}{2}$ modes in the NS sector. There is also some discussion about different 'polarization pictures' in the quantization of twistor space one has to be careful about.\\
     In section \ref{Intermediate} the massless limit of the model is extended by changing the auxiliary fermionic fields to an auxiliary supertwistor with additional gauging such that the total model has all anomalies cancelled with $c = 0$ and $a=1$ in \eqref{virasoro}. The spectrum is examined. When the twistor modes are in the NS sector, fixed vertex operators look like in the massless limit of the previous model, but when the modes are in the R sector, they rather resemble the ones of the Berkovits-Witten twistor string \cite{Berkovits_1:2004, Dolan:2008}, up to the little group symmetry and the absence of 'dipole' fields. It turns out that tree scattering of spin $\le1$ particles leads automatically to a Parke-Taylor factor.\\
    Motivated to simplify the vertex operators and scattering amplitudes of the model in section \ref{Intermediate}, it is extended once more in section \ref{Improved} to get a vanishing $L_0$ constant while staying anomaly-free and keeping the rich spectrum in the R sector. This is achieved mainly by reducing the bosonic auxiliary worldsheet spinors introduced in the previous section and treating the fermionic twistor modes and their duals in a more symmetric fashion. It is again a massless model without, unfortunately, a known relation to a massive version. The fixed and integrated vertex operators are determined and tree scattering amplitudes are computed, still containing a Parke-Taylor factor for spin $\le1$ particle scattering.\\
    Section \ref{Summary} contains summary and discussion, including arguments that the spectrum of the original massive model and the model of section \ref{Improved} in the R sector includes the spin 1/2 spectrum of 3 generations of the Pati-Salam model \cite{Pati:1974}, and also that the (truncated) spectrum in the NS sector can be re-interpreted as arising from spectral flow.\\   
    In appendix \ref{Appendix} the auxiliary supertwistor fields in the model of section 4 are replaced with the actual twistor fields, breaking the worldsheet supersymmetry in manifest fashion. The resulting model has the same spectrum as the section \ref{Intermediate} model, but there is no need for the little group representation to be the same across the spectrum. When adjusting the notation, this model turns out to be the same as in \cite{Kunz:2020, Kunz_1:2020}. Like the model in section \ref{Intermediate} it also has vertex operators leading to scattering amplitudes that are unappealing. This makes the model in section \ref{Improved} more favorable.\\

  \section{Notation and Oscillator Expansion}
  \label{Notation}
  The little group of the two twistor representation of a massive particle that keeps its timelike momentum $k_{\alpha \dot\alpha}$ invariant includes SU(2) as a subgroup. In the following it will be referred to as 'the little group'. Dealing in this work with complexified twistor space, SU(2) will be regarded as extended to SL(2, $\mathbb{C}$) whose algebra is the complexification of the SU(2) algebra. $k_{\alpha \dot\alpha}$ can be written using 2-spinors as 
  \begin{equation}
  \label{Momentum}
  k_{\alpha \dot\alpha} = \kappa_{a \alpha} \kappa_{\dot\alpha}^a\,,
  \end{equation}
where a = 1, 2 is an SL(2, $\mathbb{C}$) little group index raised and lowered by $\varepsilon^{ab} = \varepsilon^{[ab]}, \varepsilon_{ab} = \varepsilon_{[ab]}, \varepsilon_{12} = 1 = \varepsilon^{12}, \varepsilon_{ac} \varepsilon^{cb} = \delta^b_a$. Little group contractions are denoted by
  \begin{equation*}
  (v_1 v_2) := v_{1a} v_{2b} \varepsilon^{ab}\,.
  \end{equation*}

  The dimensionally reduced action in 4 dimensions of \cite{Albonico:2022, Albonico:2023} is
  \begin{equation}
  \label{sugra-action}
S =\int_\Sigma \mathcal{Z}^a \cdot \bar\partial \mathcal{Z}_a + A_{ab} \mathcal{Z}^a \cdot \mathcal{Z}^b + a(\lambda^2-j^H) + \tilde{a} (\tilde{\lambda}^2- j^{H}) + S_m\,,
  \end{equation}
  where
  
  \begin{itemize}
  
  \item $a = 1,2$ is the little group index and the supertwistor fields $\mathcal{Z}_a$ = $\varepsilon_{ab}\mathcal{Z}^b$ are worldsheet spinors, repackaged into \textit{Dirac} supertwistors of the form\footnote{There is some slight modification to the notation used in \cite{Albonico:2022, Albonico:2023} with regard to the position of spinor indices.}
  \begin{equation*}
    \mathcal{Z} = (\lambda_A,\mu^A,\eta^\mathcal{I})\, :\qquad \lambda_A = (\lambda_\alpha,\tilde \lambda_{\dot\alpha})\, , \quad \mu^A = (\mu^{\dot\alpha},\tilde\mu^{\alpha})\, , \quad \eta^{\mathcal{I}} = (\eta^I,\tilde \eta_I)\, ,
    \label{Dirac-twistor}
  \end{equation*}
  where $\lambda_A$ and $\mu^A$ are Dirac spinors made up of the homogeneous chiral and antichiral components of the twistor $Z = (\lambda_\alpha, \mu^{\dot\alpha})$ and dual twistor $\tilde{Z} = (\tilde \lambda_{\dot\alpha}, \tilde\mu^{\alpha})$. In the fermionic components $\eta^\mathcal{I}=(\eta^I,\tilde \eta_I)$ with $I=1,\ldots,\frac{\mathcal{N}}{2}$, the index $\mathcal{I}=1,\ldots,\mathcal{N}$ is the R-symmetry index, with $\mathcal{N}=4$ for maximal Super-Yang-Mills (SYM) and $\mathcal{N}=8$ for maximal supergravity. Indices are raised and lowered with a symmetric form for Grassmann even entities\\ 
  $\epsilon^{AB} , \epsilon_{AB}, \epsilon^{AB} \epsilon_{AC} = \delta^B_C$, for example $\lambda^A = \epsilon^{AB} \lambda_B = (\tilde \lambda^{\dot\alpha}, \lambda^\alpha)$,\\ and a skew form for Grassmann odd entities\\
  $\Omega^{IJ} , \Omega_{IJ}, \Omega^{IJ} \Omega_{IK} = \delta^J_K$, for example $\eta_I = \Omega_{IJ} \eta^J$.\\
  Also note the Dirac $\gamma_5$ matrix defined here by $\gamma_5^{AB} \rho_B = (-\rho^{\dot\alpha}, \rho^{\alpha}), \gamma_5^{AB} \gamma_{5AC} = \delta^B_C$. For Grassmann odd entities it can be used to raise and lower indices, for Grassmann even identities it defines a 'dual' version. This will become important later.\\
   The inner product is defined as\\ 
  $\mathcal{Z}_1 \cdot \mathcal{Z}_2 = \frac{1}{2}(\tilde{Z_1} \cdot Z_2 + \tilde{Z_2} \cdot Z_1 + \tilde\eta_{1I} \eta^I_2 + \tilde\eta_{2I} \eta^I_1),\;\; \tilde{Z_1} \cdot Z_2 = \tilde\mu_1^{\alpha}\lambda_{2 \alpha} + \tilde \lambda_{1 \dot\alpha} \mu_2^{\dot\alpha}$\,,\\
  with special treatment of $\bar\partial Z$ when taking the dual: $\;\widetilde{\bar\partial Z} = - \bar\partial \tilde{Z}\,$.\\
  The only non-trivial OPEs are\\
  $\mathcal{Z}^A_a(z) \cdot \mathcal{Z}_{B b}(0) = \frac{\delta^A_B \epsilon_{ab}}{z} + \cdots$\;.
  
  \item little group transformations are gauged by the fields $A_{ab} = A_{(ab)}$. $a$ and $\tilde{a}$ are worldsheet $(0,1)$-forms that act as Lagrange multipliers to constrain the mass operators
  
  $\lambda^2=\frac{1}{2}(\lambda_\alpha\lambda^\alpha)=\det(\lambda_\alpha^a)$, $\tilde{\lambda}^2=\frac{1}{2}(\tilde{\lambda}_{\dot\alpha} \tilde{\lambda}^{\dot\alpha})=\det(\tilde{\lambda}_{\dot\alpha}^a)$
  
  to be the same as the current $j^H$ associated to the element $h \in \mathcal{G}$ living in the Cartan subalgebra of some symmetry of the system. This article will not be concerned about the particularity of particle masses. Therefore, there will be no further discussion of $j^H$.
  
  \item the action $S_m$ represents worldsheet matter and different choices for $S_m$ construct a variety of physically interesting models. For supergravity $S_m$ is equal to $S_{\rho_1} + S_{\tilde\rho_2}$ with 
  \begin{equation*}
  \label{action-WS-fermions}
  S_{\rho}=\int_{\Sigma} \tilde{\rho}^{A} \bar{\partial} \rho_{A}+b_{a}\lambda^{Aa} \rho_{A}+\tilde{b}_{a} \lambda_{A}^{a} \tilde{\rho}^{A}\,,
  \end{equation*} 
  where $(\rho_A,\tilde\rho^A)$ are fermionic worldsheet spinors raised and lowered with $\gamma_5$ and $(b_{a}, \tilde{b}_{a})$ are $(0,1)$-forms on the worldsheet acting as fermionic Lagrange multipliers for the constraints $\lambda^{Aa} \rho_{A}=0=\lambda_{A}^{a} \tilde{\rho}^{A}$.
  
  \end{itemize}
  
  In ambitwistor space vertex operators are typically built with help of plane wave representatives. For the current supersymmetric system this looks like
  \begin{align}
  \label{VO}
  &\mathcal{V} = \qquad \int \!\! d^{2} u\; d^{2} v\;\; \mathcal{W}(u) \;\; \bar{\delta}^{4}\!\left(\left(u \lambda_{A}\right)-\left(v \kappa_{A}\right)\right)\; \bar{\delta}((\epsilon v)-1) e^{u_{a}\left(\mu^{A a} \epsilon_{A}+\eta^{\mathcal{I} a}q_{\mathcal{I}} \right)-\frac{1}{2}(\xi v) q^{2}}\;,\nonumber\\
  &V= \int_\Sigma \!\!d\sigma \!\! \int \!\! d^{2} u\; d^{2} v\;\; w(u) \;\; \bar{\delta}^{4}\!\left(\left(u \lambda_{A}\right)-\left(v \kappa_{A}\right)\right)\; \bar{\delta}((\epsilon v)-1) e^{u_{a}\left(\mu^{A a} \epsilon_{A}+\eta^{\mathcal{I} a}q_{\mathcal{I}} \right)-\frac{1}{2}(\xi v) q^{2}}\;,
  \end{align}
  where $\mathcal{V}$ and $V$ stand for a fixed and integrated vertex operator, respectively,
  polarization data $\epsilon_A$ is defined by $\epsilon_A = \epsilon_a \kappa^a_A$ with $\kappa^a_A = (\kappa^a_\alpha, \kappa^a_{\dot \alpha})$ being the momentum of the wave according to \eqref{Momentum},
  super polarization data $q_{\mathcal{I}}$ is defined by $q_{\mathcal{I}} = \epsilon_a q^a_{\mathcal{I}}$ with $q^a_{\mathcal{I}}$ being the supermomentum, and
  $(\epsilon_a, \xi_a)$ with $(\epsilon \xi) = 1$ form a basis of the fundamental representation of SL(2, $\mathbb{C}$) so that the supersymmetry generators can be defined as $Q_{a \mathcal{I}} = (\xi_a q_{\mathcal{I}} + \epsilon_a \Omega_{\mathcal{I} \mathcal{J}} \frac{\partial}{\partial q_{\mathcal{J}}})$\cite{Albonico:2020}. \\
  
  The quadratic differentials $\mathcal{W}$ and $w$ in \eqref{VO} are theory dependent and are allowed to depend on the parameter $u$ as well as the worldsheet matter systems. For a fixed vertex operator in supergravity $\mathcal{W}$ is just the product of fermionic ghost fields and delta functions of bosonic ghosts. For an integrated vertex operator an additional integration over the worldsheet is applied in \eqref{VO} to take care of gauge fixing the worldsheet diffeomorphisms and $w$ is \cite{Albonico:2022}
  \begin{align}
  \label{vo-matter}
    w(u) =\;&
    \delta\big( \mathrm{Res}_\sigma (\lambda^2-j^H)\big) \; \delta\big(\mathrm{Res}_\sigma (\tilde\lambda^2-j^H)\big)\nonumber\\
    &\left(\frac{(\hat{u} \lambda_{A}) \epsilon^{A}}{(u \hat{u})}-\epsilon^{A} \epsilon_{B} \rho_{1A} \tilde{\rho}_1^{B}\right)\,
    \left(\frac{(\hat{u} \lambda_{A}) \tilde{\epsilon}^{A}}{(u \hat{u})}-\tilde{\epsilon}^{A} \tilde{\epsilon}_{B} \rho_{2A} \tilde{\rho}_2^{B}\right)\,,
  \end{align}
  where $\hat{u}$ is chosen such that $(u \hat{u}) \ne 0$ \footnote{Under the support of the polarized scattering equations $w(u)$ will not depend on the particular value of $\hat{u}$\cite{Geyer:2020}.} and where the dual polarization data $\tilde{\epsilon}^A$ is defined with help of the $\gamma_5$ matrix: $\tilde{\epsilon}^A = \gamma_5^{AB} \epsilon_B$.\\
    
    According to standard BRST procedure, in addition to the familiar fermionic $(b,c)$ ghosts related to worldsheet gravity (the action \eqref{sugra-action} is already written in conformal gauge), the Lagrange multipliers in \eqref{sugra-action} are associated with corresponding ghosts:\\
    the bosonic fields $\{A_{a}, a, \tilde{a}\}$ with fermionic ghosts $\{(M_{ab}, N_{ab}), (m, n), (\tilde{m}, \tilde{n})\}$ and\\
    the fermionic fields $\{b^a_r, \tilde{b}^a_r\}$ with bosonic ghosts $\{(\beta^a_r, \gamma^a_r),(\tilde{\beta}^a_r, \tilde{\gamma}^a_r)\}$.\\
    
    The SL(2,$\mathbb{C}$) anomaly coefficient is:
    \begin{equation*}
    a_{sl2} = \frac{3}{2}(\frac{1}{2}(8 - \mathcal{N}))_{\mathcal{Z}} + \frac{3}{2}(2_{\beta \gamma} + 2_{\tilde{\beta} \tilde{\gamma}}) - 6_{M\!N} = \frac{3}{4}(8 - \mathcal{N})\,.
    \end{equation*}
    It vanishes for $\mathcal{N} = 8$.\\
    The central charge is 
    \begin{equation*}
    c = (-8 + \mathcal{N})_{\mathcal{Z}} - 26_{bc} - 6_{M\!N} - 2_{mn} - 2_{\tilde{m}\tilde{n}} + 2(4_{\rho \tilde{\rho}} + 4_{\beta \gamma} + 4_{\tilde{\beta} \tilde{\gamma}}) = -20 + \mathcal{N}\,.
    \end{equation*}
    It vanishes again for $\mathcal{N} = 8$ when a remaining central charge of $c = -12$ is accounted for from compactifying 6 dimensions with a central charge $c_{6d} = 12$.\\
    
    Setting $\mathcal{N} = 8$ from now on, the $L_0$ constant $a$ is given by
    \begin{equation*}
    24 a = 0_{\mathcal{Z}} - 2_{bc} - 6_{M\!N} - 2_{mn} - 2_{\tilde{m}\tilde{n}} + 2(a_{\rho \tilde{\rho}} + 4_{\beta \gamma} + 4_{\tilde{\beta} \tilde{\gamma}}) = 4 + 2 a_{\rho \tilde{\rho}} + a_{6d}\,,
    \end{equation*}
    where $a_{\rho \tilde{\rho}} = - 8$ in the R sector and $a_{\rho \tilde{\rho}} = 4$ in the NS sector, i.e. $a = - \frac{1}{2} + \frac{1}{24}a_{6d}$ in the R sector and $a = \frac{1}{2} + \frac{1}{24}a_{6d}$ in the NS sector.\\
    
    The actual spectrum of the model can be determined by considering oscillator expansions of the supertwistor fields, given here on the Riemann sphere:
    \begin{equation}
    \label{Twistor-Expansion}
  \lambda_{aA} = \sum_n \! \lambda_{aA n} \sigma^{-n-\frac{1}{2}}, \quad \mu_a^A  = \sum_n \! \mu_{a n}^A \sigma^{-n-\frac{1}{2}},\quad  
  \eta_a^{\mathcal{I}} = \sum_n \! \eta_{an}^{\mathcal{I}} \sigma^{-n-\frac{1}{2}}\,,
  \end{equation}
  with $n \, \epsilon \, \mathbb{Z}$ in the R sector and $n \, \epsilon \, \mathbb{Z} + \frac{1}{2}$ in the NS sector.
  The expansions of the energy momentum $T$ and its zero mode $L_0$ are after normal ordering
  \begin{equation}
  \label{L0-Expansion}
  T= \sum_{n \in \mathbb{Z}}  \! L_n \sigma^{-n-2}, \quad L_0  = \!\sum_{n \in \mathbb{Z}} \! n\, :\mu_{a n}^A \lambda^a_{A-n}: + \sum_{n \in \mathbb{Z}} \! n\, :\tilde{\eta}^a_{I-n} \,\eta^I_{an}: + \ldots \,,
  \end{equation} 
  where $\dots$ denotes contributions from non-twistor fields.
    \\

\section{Spectrum of the Supergravity Model}
  \label{Spectrum}
  
  After reviewing the notation in the previous section, the spectrum can now be analyzed. It is clear from looking at the fixed vertex operator in \eqref{VO} and the expansion of $L_0$ in \eqref{L0-Expansion} that the spectrum is either generated by appropriate homogeneous functions of just zero modes of $\lambda_A^a$ and $\eta^{\mathcal{I}}_a$ in the R sector or by products of exactly two $-\frac{1}{2}$ modes of $\lambda_A^a$ and $\eta^{\mathcal{I}}_a$ in the NS sector, otherwise the vertex operator would be required to show derivatives of the twistor fields\footnote{For instance such vertex operators appear for the Berkovits-Witten twistor string \cite{Dolan:2008}.}. The $\mu^A_{an}$ modes are excluded from the spectrum because BRST cohomology requires that the non-negative modes of the $\lambda^{Aa}\rho_A$ current annihilate physical states.\\
  
  Therefore, the model must have an $L_0$ constant that is $a = 0$ with $a_{6d} = 12$ in the R sector and $a = 1$ with the same $a_{6d} = 12$ in the NS sector. But there is also the need to give a central charge contribution of exactly 12. The requirement of $c_{6d} = 12 = a_{6d}$ can easily be achieved by adding 6 bosonic scalars (see appendix A in \cite{Albonico:2023}). On a side note, in the NS sector, $a_{6d}$ = $c_{6d}$ always holds independently of the number of fermionic or bosonic worldsheet scalar or spinor fields added for compactification, i.e. always $a = 1$ in the NS sector. This is not true for the R sector.\\
  
  The conclusion is that the model of \cite{Albonico:2023} operates in both the R and NS sector, using the same vertex operators \eqref{VO}. One important remark concerning the lowest modes of bosonic twistor components needs to be added. When taking the massless limit, the vertex operator \eqref{VO} can be made to degenerate into a vertex operator of positive helicity and one of negative helicity \cite{Albonico:2020, Albonico:2023}. A vertex operator for positive helicity uses the twistor components $\lambda^a_{\alpha 0}$ or $\lambda^a_{\alpha -\frac{1}{2}}$ as creation operators, and a vertex operator for negative helicity uses the dual twistor components $\tilde{\lambda}^a_{\dot\alpha 0}$ or $\tilde{\lambda}^a_{\alpha -\frac{1}{2}}$ as creation operators. But they cannot co-exist in the same picture, otherwise a combination of them could be used to generate physical states without being able to assign a helicity (the 'googly' problem). This is similar to being unable to choose a K\"{a}hler polarization simultaneously on both twistor and dual twistor space \cite{Woodhouse:1975jb} (see also Table \ref{RSpectrumTable} in next section).\\
  
  In other words, the vertex operator \eqref{VO} hides the fact of potentially operating in two different pictures. In the R sector one picture has the first two components in $\lambda^a_{A0}$ and $\mu^{aA}_0$ as creation operators and the last two as annihilation operators and the other picture has the nature of the components reversed. In the NS sector it is even more complicated: in one picture $\tilde{\lambda}^a_{\dot{\alpha}-\frac{1}{2}}$ is an annihilation operator interchanging with $\mu^{a \dot\alpha}_{\frac{1}{2}}$ as creation operator and in the other picture $\lambda^a_{{\alpha}-\frac{1}{2}}$ is an annihilation operator swapped with $\tilde{\mu}^{a \alpha}_{\frac{1}{2}}$ as creation operator\footnote{This is a well-known fact already observed for the original 4-dimensional ambitwistor\cite{Lipstein:2015}. Note that, depending on the polarization, the index of some oscillators in \eqref{Twistor-Expansion} might get shifted by 1.}. The scattering equations seem to be able to interpolate between the two pictures. This becomes more evident in the massless limit and the next sections will focus more on this limit, although in section \ref{Improved} it will be seen that with maximal supersymmetry the vertex operators in the massive model can stay in one picture and still generate the full $\mathcal{N} \!=\! 8$ supergravity particle spectrum, i.e. there is no inconsistency in using the vertex operators in \eqref{VO}. For the remainder of the article the two pictures are referred to as 'polarization pictures'.\\
  
  One additional observation in the NS sector is that only two $-\frac{1}{2}$ modes can appear in the spectrum in order for $L_0 = 1$ to be valid, i.e. the supersymmetric spectrum is truncated and only contains a single graviton and some spin $\frac{3}{2}$ and spin 1 particles but no scalars or spin $\frac{1}{2}$ fermions. This also implies that all particles have to remain massless in the NS sector. This issue will be discussed more, with a possible solution, in the last section \ref{Summary} reserved for summary and discussion.\\

\section{Intermediate Modification of the Supergravity Model}
  \label{Intermediate}
  If one is on the lookout for a self-contained supergravity model with a spectrum that includes particles that can be interpreted as gluons or quarks, one requirement would be that the scattering of such particles leads to a Parke-Taylor factor. The model in the previous sections does not fulfill this condition without an external current algebra. This section introduces a modification that keeps the model anomaly-free but allows for such a Parke-Taylor factor. On the other hand, the model picks up some undesired features. In the next section it will get an additional extension that makes it more satisfactory. Nevertheless, it is worthwhile to not skip the intermediate step because it relates closely to other models in the literature.\\
  
  The modification of the model consists by not fixing the masses of the two twistors and insisting on massless particles only, by doubling the number of fermionic components of the supertwistor, and by gauging them in similar fashion to the bosonic components with help of auxiliary fields:  
  \begin{equation*}
  \label{intermediate-sugra-action}
S =\int_\Sigma \mathcal{Z}^a \cdot \bar\partial \mathcal{Z}_a + A_{ab} \mathcal{Z}^a \cdot \mathcal{Z}^b + S_{\rho_1} + S_{\tilde{\rho}_2} + S_{\tau_1} + S_{\tau_2}\,,
  \end{equation*}
  with
  \begin{align*}
  %\label{action-WS-bosons} 
  &S_{\rho}=\int_{\Sigma} \tilde{\rho}^A \bar{\partial} \rho_A + b_a \lambda^{aA} \rho_A  + \tilde{b}_{a} \lambda^a_A \tilde{\rho}^A\,,\nonumber\\  
  &S_{\tau_1}=\int_{\Sigma} \tilde{\tau}_1^{\mathcal{I}} \bar{\partial} \tau_{1\mathcal{I}} + d_{1a}\eta^{a I} \tau_{1I} + \tilde{d}_{1a} \eta_I^a \tilde{\tau}_1^I\,,\\ 
  &S_{\tau_2}=\int_{\Sigma} \tilde{\tau}_2^{\mathcal{I}} \bar{\partial} \tau_{2\mathcal{I}} + d_{2a}\tilde{\eta}_I^a \tau_2^{\prime I} + \tilde{d}_{2a} \tilde{\eta}^{a I}\tilde{\tau}^{\prime}_{2I}\,,\nonumber
  \end{align*} 
  where supertwistor fields $\mathcal{Z}$ have been extended to 
  \begin{align*}
  \label{supertwistor}
    \mathcal{Z}=(\lambda_A,\mu^A,\eta^\iota)\, :\quad &\lambda_A=(\lambda_\alpha,\tilde \lambda_{\dot\alpha})\, ,  \quad \mu^A=(\mu^{\dot\alpha},\tilde\mu^{\alpha})\, ,\quad \eta^{\iota}=(\eta^\mathcal{I}, \tilde{\eta}_\mathcal{I})\,,\;\eta^{\mathcal{I}}=(\eta^I,\tilde \eta_I)\, , \;\\
    &\tilde{\eta}_\mathcal{I}=(\tilde \eta^\prime_I, \eta^{\prime I})\, , \;I = 1 \ldots \frac{\mathcal{N}}{2} , \,\mathcal{I} = 1, \ldots, \mathcal{N}, \,\iota = 1, \ldots, 2\mathcal{N}\,, \,\mathcal{N} = 8\, ,\nonumber
  \end{align*} 
  $(\tau_\mathcal{I},\tilde\tau^\mathcal{I}) = ((\tau_I, \tau^{\prime I}), (\tilde{\tau}^I, \tilde{\tau}^\prime_I))$ are bosonic worldsheet spinors,
  and $(b_{ra}, \tilde{b}_{ra}), (d_{ra}, \tilde{d}_{ra})$ are $(0,1)$-forms on the worldsheet acting as fermionic Lagrange multipliers for the constraints\\
  $\lambda^{aA} \rho_{1A} =  \lambda^a_A \tilde{\rho}_1^A = \lambda^{aA} \tilde{\rho}_{2A} =  \lambda^a_A \rho_2^A = \eta^{a I} \tau_{1I} = \eta_{I}^{a} \tilde{\tau}_{1}^{I} = \tilde{\eta}_I^{a} \tau_{2}^{\prime I} = \tilde{\eta}^{a I}\tilde{\tau}^{\prime}_{2I} = 0$,\\
  
  During BRST quantization the additional fermionic fields $\{d_{ra}, \tilde{d}_{ra}\}$ lead to new bosonic ghosts $\{(\beta_{ra}^{\prime}, \gamma_{ra}^{\prime}),(\tilde{\beta}^{\prime}_{ra}, \tilde{\gamma}^{\prime}_{ra})\}$.
  The SL(2,$\mathbb{C}$) anomaly coefficient becomes:
    \begin{equation*}
    a_{sl2} = \frac{3}{2}(\frac{1}{2}(8 - 2\mathcal{N}))_{\mathcal{Z}} + \frac{3}{2}(2_{\beta \gamma} + 2_{\tilde{\beta} \tilde{\gamma}} + 2_{\beta^\prime \gamma^\prime} + 2_{\tilde{\beta}^\prime \tilde{\gamma}^\prime}) - 6_{M\! N} = \frac{3}{2}(8 - \mathcal{N}) = 0\,.
    \end{equation*}
    It is still zero. The central charge vanishes as well:
    \begin{equation*}
    c = (-8 + 2 \mathcal{N})_{\mathcal{Z}} - 26_{bc} - 6_{{M\! N}} + 2(4_{\rho \tilde{\rho}} + 4_{\beta \gamma} + 4_{\tilde{\beta} \tilde{\gamma}}) + 2(-\mathcal{N}_{\tau \tilde{\tau}} + 4_{\beta^\prime \gamma^\prime} + 4_{\tilde{\beta^\prime} \tilde{\gamma^\prime}}) = 0\,.
    \end{equation*}
    As mentioned earlier, the $L_0$ constant $a$ in the NS sector does not change and stays equal to 1 and in the R sector:
    \begin{equation*}
    24 a = (16 - 4\mathcal{N})_{\mathcal{Z}} - 2_{bc} - 6_{{M\! N}} + 2(-8_{\rho \tilde{\rho}} + 4_{\beta \gamma} + 4_{\tilde{\beta} \tilde{\gamma}}) + 2(2\mathcal{N} _{\tau \tilde{\tau}} + 4_{\beta^\prime \gamma^\prime} + 4_{\tilde{\beta}^\prime \tilde{\gamma}^\prime}) = 24\,.
    \end{equation*}
    i.e. $a = 1$ in the R sector as well.\\
    
  Although the fermionic modes in the supertwistors are doubled, similarly to the exclusion of $\mu^A_n$ modes from the spectrum because of BRST cohomology, now the $\tilde{\eta}_{\mathcal{I} n}$ cannot contribute to the spectrum because of the requirement that the non-negative modes of the $\eta^{a I} \tau_{1 I}$ and $\tilde{\eta}^a_I \tau_2^{\prime I}$ currents annihilate physical states.
  This leaves the spectrum in the NS sector unchanged but the one in the R sector gets modified considerably, because there always has to be a single $-\!1$ twistor mode of $\lambda_A^a$ or $\eta^{\mathcal{I}}_a$ in addition to an appropriate homogeneous function of just zero modes.\\
    
 The internal little group representation is assumed to be in a singlet, making sure that there is only one graviton-like excitation. The R-symmetry is then an SU(4) group in each of the two pictures, one with polarization $\tilde{\lambda}_{\dot\alpha} \!\sim\! \partial/\partial\mu^{\dot\alpha}, \tilde{\mu}^\alpha \!\sim\! \partial/\partial\lambda_\alpha, \tilde{\eta}_I \!\sim\! \partial/\partial\eta^{\prime I}, \tilde{\eta}_I^\prime \!\sim\! \partial/\partial\eta^I$ for the zero modes, the other one with polarization $\lambda_\alpha \!\sim\! \partial/\partial\tilde{\mu}^\alpha, \mu^{\dot\alpha} \!\sim\! \partial/\partial\tilde{\lambda}_{\dot\alpha}, \eta^I \!\sim\! \partial/\partial\tilde{\eta}_I^\prime, \eta^{\prime I} \!\sim\! \partial/\partial\tilde{\eta}_I$. Table \ref{RSpectrumTable} displays the spectrum, using the standard notation
   \begin{equation*}
    \braket{\lambda g} = \varepsilon^{\alpha \beta} \lambda_{\alpha} g_{\beta} = -\braket{g \lambda} \qquad
    [\tilde\lambda f] = \varepsilon^{\dot\alpha \dot\beta} \tilde{\lambda}_{\dot\alpha} f_{\dot\beta} = -[f \tilde\lambda]\,.
\end{equation*}
All helicity states have double occurrence, reflecting the symmetry between twistor fields and their duals, and being related to each other through a Fourier transformation \cite{Berkovits_1:2004}. The table also shows a strong resemblance with part of the conformal supergravity spectrum of the Berkovits-Witten twistor string with N=4 supersymmetry \cite{Berkovits_1:2004, Dolan:2008}. However, because of the supersymmetric gauging it does not contain the 'dipole' states thought of being responsible for the lack of unitarity in the Berkovits-Witten model \cite{Berkovits_1:2004}.\\
    
 {\renewcommand{\arraystretch}{2.0}
 \begin{table}[hbt!]
 \begin{center}
 \begin{tabular}{||c|c||} 
 \hline
 Oscillators & (Helicity| SU(4)) \\ [0.5ex] 
 \hline\hline
 %&\\
 $[\tilde{\lambda}_{-\!1}^{a} f_a(\lambda^b_{\alpha 0},\!\eta^{I}_{c 0})]$ & $(2|1), \,\,\,\,\,\,\,\,\,(\frac{3}{2}|\bar{4}), \,\,\,\,\,\,\,\,\,(1|6), \,\,\,\,\,\,\,\,\,(\frac{1}{2}|4), \,\,\,\,\,\,\,\,\,(0|1)$\\ 
 %&\\
 \hline
 %&\\
 $\braket{\lambda^a_{-\!1} g_a(\lambda^b_{\alpha 0},\!\eta^{I}_{c 0})}$ & $(0|1), \,\,\,\,\,\,(-\!\frac{1}{2}|\bar{4}), \,\,\,\,\,(-\!1|6), \,\,\,\,\,(-\!\frac{3}{2}|4), \,\,\,\,\,(-\!2|1)\!$ \\ 
 %&\\
 \hline
 %&\\
 $\tilde{\eta}^a_{I -\!1} f^I_a(\lambda^b_{\alpha 0},\!\eta^{I}_{c 0})$ & $(\frac{3}{2}|4), \,(1|4 \!\otimes\! \bar{4}), \,\,(\frac{1}{2}|4 \!\otimes\!  6), \,\,(0|4\!\otimes\!  4), \,\,(-\!\frac{1}{2}| 4)$ \\
 %&\\
 \hline
 %&\\
 $\eta^I_{a -\!1} g_I^a(\lambda^b_{\alpha 0},\!\eta^{I}_{c 0})$ & $(\frac{1}{2}|\bar{4}), (0|\bar{4} \!\otimes\! \bar{4}), (-\!\frac{1}{2}|\bar{4} \!\otimes\!  6), (-\!1|\bar{4} \!\otimes\!  4), (-\!\!\frac{3}{2}| \bar{4})$ \\
 %&\\
 \hline\hline
 %&\\
 $\braket{\lambda^a_{-\!1} \tilde{f}_a(\tilde{\lambda}^b_{\dot\alpha 0},\!\tilde{\eta}^c_{I 0})}$ & $(-\!2|1), \,\,\,(-\!\frac{3}{2}|4), \,\,\,\,\,(-\!1|6), \,\,\,\,\,\,(-\!\frac{1}{2}|\bar{4}), \,\,\,\,\,\,\,(0|1)$ \\
 %&\\
 \hline
 %&\\
 $[\tilde{\lambda}^a_{-\!1} \tilde{g}_a(\tilde{\lambda}^b_{\dot\alpha 0},\!\tilde{\eta}^c_{I 0})]$ & $\,(0|1), \,\,\,\,\,\,\,\,\,(\frac{1}{2}|4), \,\,\,\,\,\,\,\,\,(1|6), \,\,\,\,\,\,\,\,\,(\frac{3}{2}|\bar{4}), \,\,\,\,\,\,\,\,\,(2|1)$ \\
 %&\\
 \hline
 %&\\
 $\eta^{I}_{a -\!1} \tilde{f}_I^a(\tilde{\lambda}^b_{\dot\alpha 0},\!\tilde{\eta}^c_{I 0})$ & $(-\!\frac{3}{2}|\bar{4}), (-\!1|\bar{4} \!\otimes\! 4), (-\!\frac{1}{2}|\bar{4} \!\otimes\!  6), (0|\bar{4} \!\otimes\!  \bar{4}), (\frac{1}{2}| \bar{4})$ \\
 %&\\
\hline
 %&\\
 $\tilde{\eta}^a_{I -\!1} \tilde{g}^I_a(\tilde{\lambda}^b_{\dot\alpha 0},\!\tilde{\eta}^c_{I 0})$ &$(-\!\frac{1}{2}|4), \,\,(0|4 \!\otimes\! 4), \,\,\,\,(\frac{1}{2}|4 \!\otimes\!  6), \,\,(1|4\!\otimes\!  \bar{4}), (\frac{3}{2}| 4)$ \\
 %&\\
 \hline
 \end{tabular}
 \caption{R Spectrum. The upper half is in the picture with polarization $\tilde{\lambda}_{\dot\alpha} \!\sim\! \partial/\partial\mu^{\dot\alpha}, \tilde{\mu}^\alpha \!\sim\! \partial/\partial\lambda_\alpha, \tilde{\eta}_I \!\sim\! \partial/\partial\eta^{\prime I}, \tilde{\eta}_I^\prime \!\sim\! \partial/\partial\eta^I$ for the zero modes, the lower half in the picture with polarization $\lambda_\alpha \!\sim\! \partial/\partial\tilde{\mu}^\alpha, \mu^{\dot\alpha} \!\sim\! \partial/\partial\tilde{\lambda}_{\dot\alpha}, \eta^I \!\sim\! \partial/\partial\tilde{\eta}_I^\prime, \eta^{\prime I} \!\sim\! \partial/\partial\tilde{\eta}_I$. All helicity states have double occurrence.}
 \label{RSpectrumTable}
 \end{center} 
 \end{table}

  In the massless limit vertex operators are more conveniently distinguished by the polarization picture they are operating in, and the polarization data can be chosen in a special little group gauge such that only one of the two twistors is non-zero, different per picture, and such that the effect of one of the two $S_\rho$ actions is swallowed up during the transition from the integration measure for scattering amplitudes of the massive model to the simplified integration measure in the massless limit \cite{Albonico:2020, Albonico:2023}, thus providing vertex operators as in \cite{Geyer:2014}. Further, the two actions of auxiliary fields $S_{\tau_1}$ and $S_{\tau_2}$ can be 'distributed' among the two pictures, meaning that the $\tau_2$ fields do not contribute to vertex operators involving $\tau_1$ fields and the other way around. Then the vertex operators appear like
  \begin{align}
  \label{VO_1}
  \mathcal{V} &= \qquad \int \!\frac{du}{u^3} \;\; \mathcal{W}(u) \; \bar{\delta}^{2}\!\left(u \lambda_{\alpha}-\epsilon_{\alpha}\,\right) \; e^{u \left(\mu^{\dot\alpha} \tilde{\epsilon}_{\dot\alpha}+\eta^{\prime I} q_I+ \eta^{I} q^\prime_I \right)}\;,\nonumber\\
  \tilde{\mathcal{V}} &=  \qquad \int \!\frac{du}{u^3} \;\; \tilde{\mathcal{W}}(u) \; \bar{\delta}^{2}(u \tilde\lambda_{\dot\alpha}-\tilde\epsilon_{\dot\alpha})\;\;  e^{u \left(\tilde{\mu}^{\alpha} \epsilon_{\alpha}+\tilde{\eta}^\prime_I \tilde{q}^I + \tilde\eta_I \tilde{q}^{\prime I} \right)}\;,\nonumber\\
  V &= \int_\Sigma\!\!d\sigma \!\! \int \!\frac{du}{u^3} \;\;\, w(u) \;\; \bar{\delta}^{2}\!\left(u \lambda_{\alpha}-\epsilon_{\alpha}\,\right) \; e^{u \left(\mu^{\dot\alpha} \tilde{\epsilon}_{\dot\alpha}+\eta^{\prime I} q_I+ \eta^{I} q^\prime_I \right)}\;,\\
  \tilde{V} &=  \int_\Sigma\!\!d\sigma \!\! \int \!\frac{du}{u^3} \;\;\, \tilde{w}(u) \;\; \bar{\delta}^{2}(u \tilde\lambda_{\dot\alpha}-\tilde\epsilon_{\dot\alpha})\;\;  e^{u \left(\tilde{\mu}^{\alpha} \epsilon_{\alpha}+\tilde{\eta}^\prime_I \tilde{q}^I + \tilde\eta_I \tilde{q}^{\prime I} \right)}\;,\nonumber
  \end{align}
  where for fixed vertex operators $\mathcal{W}(u)$ and $\tilde{\mathcal{W}}(u)$ are products of fermionic ghost fields and delta functions of bosonic ghost fields and in the R-sector with an additional factor of the form $\,u [\tilde{\lambda} \tilde{\epsilon}]\,$ or $\,u\, \tilde{\eta}_I \tilde{q}^I\,$ in $\mathcal{W}(u)$ and $\,u\! \braket{\lambda \epsilon}\,$ or $\,u\, \eta^I q_I\,$ in $\tilde{\mathcal{W}}(u)$, and for integrated vertex operators $w(u)$ and $\tilde{w}(u)$ are  
  \begin{align}
  \label{vo_1_massless}
    w(u) =& \; \left(u[\tilde{\lambda} \tilde{\epsilon}] - u^2 \,[\tilde{\epsilon} \rho_2]\,[\tilde\epsilon \tilde{\rho}_2]\right) \left(u\,\tilde{\eta}_I \, \tilde{q}^I - u^2 \,\,\tilde{q}_I  \tau^{\prime I}_2 \,\, \tilde{q}^J  \!\tilde{\tau}^\prime_{2J} \right)_{\!\tilde{q}^I \ne 0}\nonumber\\
   \tilde{w}(u) =& \;\left(u \!\braket{\lambda \epsilon} \!- u^2 \!\!\braket{\epsilon \rho_1}\!\braket{\epsilon \tilde{\rho}_1}\right) \left(u\,\eta^I q_I\, - u^2 \,\, q^I \tau_{1 I} \;q_J \tilde{\tau}_1^J \right)_{\!q_I \ne 0}\,,
  \end{align}
  with the same additional factor in the R sector as for fixed vertex operators. The distinctions between the $q^\prime$ and $q$ parameters and between $\tilde{q}^\prime$ and $\tilde{q}$ are actually not justified because they would lift the R-symmetry from SU(4) to SU(4) $\!\otimes\!$ SU(4) (see next section). Also, by setting $q_I \!\!=\!\! 0 (=\!\! q^\prime_I)$ and $\tilde{q}^I \!\!\!=\!\! 0 (=\!\! \tilde{q}^{\prime I})$ and omitting the new factors in \eqref{vo_1_massless} involving the $\eta$ and $\tau$ fields one obtains simpler vertex operators representing gravitons. Note that integrated vertex operators $V$ and $\tilde{V}$ containing the factor with $\tau$ fields automatically have at least one fermionic twistor component and, therefore, stand for a particle with spin$\le\! \frac{3}{2}$. Further, for every second row in Table \ref{RSpectrumTable} the corresponding integrated vertex operators are not covered by \eqref{vo_1_massless} but they are taken care of by the ones for the equivalent Fourier-transformed states in the other picture.\\
  
  Tree scattering amplitudes of these vertex operators will contain the reduced determinants of Hodges matrices like in references \cite{Geyer:2014, Geyer:2018, Albonico:2020, Geyer:2020, Albonico:2022, Albonico:2023}, and also the supersymmetric exponential factor
  \begin{equation}
  \label{super-factor}
  e^{F_\mathcal{N}} = \mathrm{exp} \!\left[ \;\sum_{\genfrac{}{}{0pt}{}{j \in -}{k \in +}} \frac{u_j u_k}{\sigma_j \!-\! \sigma_k} \tilde{q}^{I}_j q_{k I} \right]\,,\;q_I \!=\!  \Omega_{IJ} \tilde{q}^J.
  \end{equation}
  What is new here are the factors involving the $\tau$ fields. They lead to fermionic Hodges-like matrices with reduced determinants that, because of the Grassmann odd nature of the components of the super polarization data, are limited in the sense that the same location can occur at most 4 times per term, and if one assumes that all vertex operators with $\tau$ fields stand for particles with spin 1 then each term reduces to a product of propagators between all locations of the same helicity, each end of the propagators appearing exactly twice up to two showing up only once, one of them representing the location of a fixed vertex operator. By pulling down twice the exponent from \eqref{super-factor}, one for each fixed vertex operator, one can connect two propagator products, one for positive and one for negative helicities, and arrive at a Parke-Taylor factor for all these particles. More details will be given in the next section for the more interesting improved model.\\
  
  The current model has some undesired features in the R sector. The vertex operators have these additional factors arising from non-zero twistor modes, which make the scattering amplitudes look unattractive. Also, although the model features a Parke-Taylor factor for scattering of spin 1 particles, it does not allow for full permutation symmetry of the entries (the same helicities need to be arranged together).\\
  
  In appendix \ref{Appendix} another model found by the author earlier is described in the current notation which has the same spectrum as in Table \ref{RSpectrumTable} with similarly unappealing scattering amplitudes.\\

\section{Improved Anomaly-free Supergravity Model Extension} 
  \label{Improved}
  The model of the previous section could be improved by making the $L_0$ constant $a$ zero in the R sector. This can be done by introducing 4 bosonic Lagrange multipliers with associated fermionic ghost-antighost pairs and adding 8 fermionic worldsheet spinors or, equivalently, reducing 8 bosonic worldsheet spinors, which keeps the central charge unchanged but reduces $a$ by 1.\\
  
  The minimally changed model is:
  \begin{align}
  \label{improved-massless-sugra-action}
S = \int_\Sigma &\mathcal{Z}^a \cdot \bar\partial \mathcal{Z}_a + A_{ab} \mathcal{Z}^a \cdot \mathcal{Z}^b + a \lambda^2 + \tilde{a} \tilde{\lambda}^2 + S_{\rho_1} + S_{\tilde{\rho}_2} + S_{\tau_1} + S_{\tau_2}\,,\nonumber\\ 
&\mathcal{Z}=(\lambda_A,\mu^A,\eta^\iota)\,, \; \eta^{\iota}=(\eta^\mathcal{I}, \tilde{\eta}_\mathcal{I})\,,\;\eta^{\mathcal{I}}=(\eta^I,\tilde \eta_{\dot{I}})\, , \;\tilde{\eta}_\mathcal{I}=(\tilde \eta^\prime_I, \eta^{\prime \dot{I}})\, ,\nonumber \\
&I,\dot{I} = 1 \ldots \frac{\mathcal{N}}{2} , \,\mathcal{I} = 1, \ldots, \mathcal{N}, \,\iota = 1, \ldots, 2\mathcal{N}\,, \,\mathcal{N} = 8\, ,\\  
  S_{\rho}=\int_{\Sigma} & \tilde{\rho}^A \bar{\partial} \rho_A + b_a \lambda^{aA} \rho_A  + \tilde{b}_{a} \lambda^a_A \tilde{\rho}^A\,,\nonumber\\ 
  S_{\tau_1}=\int_{\Sigma} & \tilde{\tau}_1^I  \, \bar{\partial} \tau_{1 I} + d_{1a} \eta^{a I} \tau_{1I} + \tilde{d}_{1a} \eta^a_I \tilde{\tau}_1^I + g_1 \tilde{\tau}_1^I \tau_{1 I}\,,\nonumber\\
  S_{\tau_2}=\int_{\Sigma} & \tilde{\tau}_{2 \dot{I}}\, \bar{\partial} \tau_2^{\dot{I}} + d_{2a} \tilde{\eta}^a_{\dot{I}} \tau_2^{\dot{I}} + \tilde{d}_{2a} \tilde{\eta}^{a \dot{I}} \tilde{\tau}_{2 \dot{I}} + g_2 \tilde{\tau}_{2 \dot{I}} \tau_2^{\dot{I}}\,,\nonumber
  \end{align} 
  where $\mathcal{Z}$ is the extended supertwistor from last section, $(\rho_A, \tilde\rho^A)$ are the same auxiliary fermionic worldsheet spinors, $\tau_{1 I}$ and $\tau_{2 \dot{I}}$ are auxiliary bosonic worldsheet spinors gauging the extended fermionic components of the supertwistor, $(d_{ra}, \tilde{d}_{ra})$ and $(g_r)$ are fermionic and bosonic Lagrange multipliers, respectively, for gauge constraints satisfied by the $\tau$ fields, and $a$ and $\tilde{a}$ are the Lagrange multipliers from the original action \eqref{sugra-action} in section \ref{Notation}. $j^H$ is set to 0 here to ensure the massless limit, mainly because integrated vertex operators for the massive case are unknown. Also, the second half $\dot{I}$ of indices $\mathcal{I}$ and the indices of $\tau_{2 \dot{I}}$ have been dotted, for later convenience.\\
  
  One observation is that the $\rho_r$ and $\tau_r$ fields could be bundled into a couple of supertwistors with reversed statistics, as has been done in Skinner's model \cite{Skinner:2013}, although there they transform under the little group. They do not contribute to the spectrum and are considered purely as auxiliary, not as matter fields.\\
  
  During BRST quantization the additional fermionic $\{d_{ra}, \tilde{d}_{ra}\}$ and bosonic $\{g_r\}$ fields lead to other new bosonic $\{(\beta^{\prime}_{ra}, \gamma^{\prime}_{ra}),(\tilde{\beta}^{\prime}_{ra}, \tilde{\gamma}^{\prime}_{ra})\}$ and fermionic ghosts $\{(s_r, t_r))\}$, respectively. The SL(2,$\mathbb{C}$) anomaly coefficient then becomes like in section \ref{Intermediate}:
    \begin{equation*}
    a_{sl2} = \frac{3}{2}(\frac{1}{2}(8 - 2\mathcal{N}))_{\!\mathcal{Z}} - 6_{M\! N} + \frac{3}{2}(2_{\beta \gamma} + 2_{\tilde{\beta} \tilde{\gamma}} + 2_{\beta^\prime \gamma^\prime} + 2_{\tilde{\beta}^\prime \tilde{\gamma}^\prime} ) = \frac{3}{2}(8 - \mathcal{N}) = 0\,.
    \end{equation*}
    The central charge is:
    \begin{equation*}
    c \!=\! (\!-8 + 2 \mathcal{N}\!)_{\!\mathcal{Z}} \!- 26_{bc} \!- 6_{M\! N} \!-2_{mn} \!- 2_{\tilde{m}\tilde{n}} \!+ 2(4_{\rho} + 4_{\beta \gamma} + 4_{\tilde{\beta} \tilde{\gamma}}\! \!-\! \frac{1}{2}\mathcal{N}_{\!\tau} \!+ 4_{\beta^\prime \gamma^\prime} \!+ 4_{\tilde{\beta^\prime} \tilde{\gamma^\prime}}\! \!- 2_{st}) \!=\! -8 + \! \mathcal{N} \!=\! 0\,.
    \end{equation*}
    The $L_0$ constant $a$ stays 1 in the NS sector and in the R sector it changes to:
    \begin{equation*}
    24 a \!=\! \!(\!16 - \!4\mathcal{N}\!)_{\!\mathcal{Z}} \!-\! 2_{bc} \!- 6_{M\! N} \!-2_{mn} \!- 2_{\tilde{m}\tilde{n}} \!+ 2(\!-8_{\!\rho} \!+ 4_{\beta \gamma} \!+ 4_{\tilde{\beta} \tilde{\gamma}} \!+ \! \mathcal{N} _{\!\tau}  \!+ 4_{\beta^\prime \gamma^\prime} \!+ 4_{\tilde{\beta^\prime} \tilde{\gamma^\prime}}\! \!- 2_{st}) \!=\! 16 - \!2 \mathcal{N} \!=\! 0\,.
    \end{equation*}
    i.e. $a = 0$ in the R sector, as desired. And the model is anomaly-free as well.\\
    
 \begin{table}[hbt!]
 \begin{center}
 \begin{tabular}{||c|c|c|c|c|c||} 
 \hline\hline
 %Fermionic &&&&&\\ 
 Oscillators & & $\eta^I$ & $\eta^I_1 \eta^I_2$ & $\eta^I_1 \eta^I_2 \eta^I_3$ & $\eta^I_1 \eta^I_2 \eta^I_3 \eta^I_4$ \\  [0.5ex] 
 %&&&&&\\
 \hline\hline
 %&&&&&\\
 & $(2|1 \!\otimes\! 1)$ & $\;\,(\frac{3}{2}|1 \!\otimes\! \bar{4})$ & $\;\,(1|1 \!\otimes\! 6)$ & $\;\,(\frac{1}{2}|1 \!\otimes\! 4)$ & $\;\;(0|1 \!\otimes\! 1)$ \\ 
 %&&&&&\\
 \hline
 %&&&&&\\
 $\tilde{\eta}_{\dot{I}}$ & $(\frac{3}{2}|4 \!\otimes\! 1)$ & $\;\,(1|4 \!\otimes\! \bar{4})$ & $(\frac{1}{2}|4 \!\otimes\!  6)$ & $\;\,(0|4\!\otimes\!  4)$ & $(-\!\frac{1}{2}| 4 \!\otimes\! 1)$\\ 
 %&&&&&\\
 \hline
 %&&&&&\\
 $\tilde{\eta}_{1 \dot{I}} \tilde{\eta}_{2 \dot{I}}$ & $(1|6 \!\otimes\! 1)$ & $\;\,(\frac{1}{2}|6 \!\otimes\!  \bar{4})$ & $(0|6\!\otimes\!  6)$ & $(-\!\frac{1}{2}|6 \!\otimes\!  4)$ & $(-\!1| 6 \!\otimes\! 1)$\\ 
 %&&&&&\\
 \hline
 %&&&&&\\
 $\tilde{\eta}_{1 \dot{I}} \tilde{\eta}_{2 \dot{I}} \tilde{\eta}_{3 \dot{I}}$ & $(\frac{1}{2}|\bar{4} \!\otimes\! 1)$ & $\;\,(0|\bar{4} \!\otimes\!  \bar{4})$ & $(-\!\frac{1}{2}|\bar{4} \!\otimes\!  6)$ & $(-\!1|\bar{4} \!\otimes\!  4)$ & $(-\!\frac{3}{2}| \bar{4} \!\otimes\! 1)$\\ 
 %&&&&&\\
 \hline
 %&&&&&\\
 $\tilde{\eta}_{1 \dot{I}} \tilde{\eta}_{2 \dot{I}} \tilde{\eta}_{3 \dot{I}} \tilde{\eta}_{4 \dot{I}}$ & $(0|1 \!\otimes\! 1)$ & $(-\!\frac{1}{2}|1 \!\otimes\! \bar{4})$ & $(-\!1|1 \!\otimes\! 6)$ & $(-\!\frac{3}{2}|1 \!\otimes\! 4)$ & $(-\!2|1 \!\otimes\! 1)$\\ 
 %&&&&&\\
 \hline
 \multicolumn{6}{c}{}\\
 \hline
 %Fermionic&&&&&\\
 Oscillators& & $\tilde{\eta}_{\dot{I}}$ & $\tilde{\eta}_{1 \dot{I}} \tilde{\eta}_{2 \dot{I}}$ & $\tilde{\eta}_{1 \dot{I}} \tilde{\eta}_{2 \dot{I}} \tilde{\eta}_{3 \dot{I}}$ & $\tilde{\eta}_{1 \dot{I}} \tilde{\eta}_{2 \dot{I}} \tilde{\eta}_{3 \dot{I}} \tilde{\eta}_{4 \dot{I}}$ \\
 %&&&&&\\
 \hline\hline
 %&&&&&\\
 & $(-\!2|1 \!\otimes\! 1)$ & $(-\!\frac{3}{2}|1 \!\otimes\! 4)$ & $(-\!1|1 \!\otimes\! 6)$ & $(-\!\frac{1}{2}|1 \!\otimes\! \bar{4})$ & $(0|1 \!\otimes\! 1)$ \\ 
 %&&&&&\\
 \hline
 %&&&&&\\
 $\eta^I$ & $(-\!\frac{3}{2}|\bar{4} \!\otimes\! 1)$ & $(-\!1|\bar{4} \!\otimes\! 4)$ & $(-\!\frac{1}{2}|\bar{4} \!\otimes\!  6)$ & $\;\,(0|\bar{4}\!\otimes\!  \bar{4})$ & $(\frac{1}{2}| \bar{4} \!\otimes\! 1)$\\ 
 %&&&&&\\
 \hline
 %&&&&&\\
 $\eta^I_1 \eta^I_2$ & $(-\!1|6 \!\otimes\! 1)$ & $(-\!\frac{1}{2}|6 \!\otimes\! 4)$ & $\;\,(0|6\!\otimes\! 6)$ & $(\;\,\frac{1}{2}|6 \!\otimes\! \bar{4})$ & $(1| 6 \!\otimes\! 1)$\\ 
 %&&&&&\\
 \hline
 %&&&&&\\
 $\eta^I_1 \eta^I_2 \eta^I_3$ & $(-\!\frac{1}{2}|4 \!\otimes\! 1)$ & $\;\,(0|4 \!\otimes\!  4)$ & $\;\,(\frac{1}{2}|4 \!\otimes\!  6)$ & $\;\,(1|4 \!\otimes\!  \bar{4})$ & $(\frac{3}{2}| 4 \!\otimes\! 1)$\\ 
 %&&&&&\\
 \hline
 %&&&&&\\
 $\eta^I_1 \eta^I_2 \eta^I_3 \eta^I_4$ & $\;\;(0|1 \!\otimes\! 1)$ & $\;\,(\frac{1}{2}|1 \!\otimes\! 4)$ & $\;\,(1|1 \!\otimes\! 6)$ & $\;\,(\frac{3}{2}| 1 \!\otimes\! \bar{4})$ & $(2| 1 \!\otimes\! 1)$\\ 
 %&&&&&\\
 \hline
 \end{tabular}
 \caption{R Spectrum. Every cell shows (helicity|SU(4) $\!\!\times\!\!$ SU(4)). The upper half is in the picture with polarization $\tilde{\lambda} \!\sim\! \partial/\partial\mu, \tilde{\mu} \!\sim\! \partial/\partial\lambda$, the lower half in the picture with polarization $\lambda \!\sim\! \partial/\partial\tilde{\mu}, \mu \!\sim\! \partial/\partial\tilde{\lambda}$. All helicity states have double occurrence of the same group representation.} 
 \label{RSpectrumTable1}
 \end{center} 
 \end{table}
 
 The spectrum can be examined from the value of the $L_0$ constant $a$. As in the previous section it might look differently depending on the selected polarization picture, $\tilde{\lambda}_{\dot\alpha} \!\sim\! \partial/\partial\mu^{\dot\alpha}, \tilde{\mu}^\alpha \!\sim\! \partial/\partial\lambda_\alpha, \tilde{\eta}_{\dot{I}} \!\sim\! \partial/\partial\eta^{\prime \dot{I}}, \tilde{\eta}_I^\prime \!\sim\! \partial/\partial\eta^I$ versus $\lambda_\alpha \!\sim\! \partial/\partial\tilde{\mu}^\alpha, \mu^{\dot\alpha} \!\sim\! \partial/\partial\tilde{\lambda}_{\dot\alpha}, \eta^I \!\sim\! \partial/\partial\tilde{\eta}_I^\prime, \eta^{\prime \dot{I}} \!\sim\! \partial/\partial\tilde{\eta}_{\dot{I}}$. Because the fermionic modes are not critical for the consistency of helicity assignment in the polarization, one can try to use in the R sector both of the zero modes $\eta^I_0$ and $\tilde{\eta}_{0 \dot{I}}$ as creation operators with $\tilde{\eta}^{\prime}_{0 I}$ and $\eta^{\prime \dot{I}}_0$ as annihilation operators in both pictures, i.e. treating them in symmetric fashion, and similarly the $-\frac{1}{2}$ modes in the NS sector. This lifts the R-symmetry from SU(4) to SU(4) $\!\!\times\!\!$ SU(4), subgroup of SU(8), as reflected in Table \ref{RSpectrumTable1}. By counting all particle modes of equal spin together one obtains the $\mathcal{N} = 8$ supergravity multiplet:\\
1 spin $\pm 2$ boson, 8 spin $\pm\frac{3}{2}$ fermions, 28 vector bosons, 56 spin $\pm\frac{1}{2}$ fermions, and 70 scalars, all in all 128 bosonic and 128 fermionic modes.\\ 

 This applies to the original massive model as well (although from a strict oscillator point of view there are only enough fermionic creation modes for $\frac{\mathcal{N}}{2} = 4$ supersymmetry). Exactly the same spectrum is covered fully in both polarization pictures, and they can be related through a Fourier transformation like in conformal supergravity\cite{Berkovits_1:2004}. Therefore, in principle one can stay just in one picture when using vertex operators and computing scattering amplitudes. This is also valid in the massive case, with some modes potentially grouped together making up massive particles, for instance\\
  $(\frac{3}{2}, -\frac{1}{2} | 4 \!\otimes\! 1), (-\frac{3}{2}, \frac{1}{2} | 1 \!\otimes\! 4), (\frac{3}{2}, -\frac{1}{2} | 1 \!\otimes\! \bar{4}), (-\frac{3}{2}, \frac{1}{2} | \bar{4} \!\otimes\! 1)$ to represent up to 8 massive spin $\frac{3}{2}$ particles or \\
  $(\pm 1 | 1 \!\otimes\! 6), (\pm 1 | 6 \!\otimes\! 1)$ and the $6$ from $(0 | 4 \!\otimes\! 4 = 10 \!\oplus\! 6), (0 | \bar{4} \!\otimes\! \bar{4} = \bar{10} \!\oplus\! 6)$ for up to 12 massive vector bosons.\\
 
 First it seems that the spectrum in the NS sector is the same as in the R sector, but because of the requirement of not having more than 2 on-shell oscillator modes it is severely truncated: the supersymmetric multiplet is reduced to one graviton with spin $\pm2$, 8 spin $\pm\frac{3}{2}$ particles, and 28 vector bosons, leaving out 56 spin $\pm\frac{1}{2}$ fermions and 70 scalars, and thus breaking target space supersymmetry. Further, without these lower spin modes, the particles cannot acquire mass through some symmetry breakdown and must remain massless. On the other hand, in the discussion of section \ref{Summary} it will be argued that the spectrum in the NS sector can be viewed as arising from the R sector through a little group gauge transformation that in the quantized theory can be interpreted as a spectral flow operation.\\
 
 Choosing the same little group gauge and polarization data as in section \ref{Intermediate} similar vertex operators \eqref{VO_1} and \eqref{vo_1_massless} are obtained in the two polarization pictures but without the additional factor in the R-sector and with the $\tau^\prime$ fields replaced by the ordinary $\tau$:
 \begin{align}
  \label{VO_2_massless}
  \mathcal{V} &= \qquad \int \!\frac{du}{u^3} \;\; \mathcal{W}(u) \; \bar{\delta}^{2}\!\left(u \lambda_{\alpha}-\epsilon_{\alpha}\,\right) \; e^{u \left(\mu^{\dot\alpha} \tilde{\epsilon}_{\dot\alpha} + \eta^{\prime \dot{I}} q_{\dot{I}} + \eta^{I} q^\prime_I \right)}\;,\nonumber\\
  \tilde{\mathcal{V}} &=  \qquad \int \!\frac{du}{u^3} \;\; \tilde{\mathcal{W}}(u) \; \bar{\delta}^{2}(u \tilde\lambda_{\dot\alpha}-\tilde\epsilon_{\dot\alpha})\;\;  e^{u \left(\tilde{\mu}^{\alpha} \epsilon_{\alpha} + \tilde{\eta}^\prime_I \tilde{q}^I + \tilde\eta_{\dot{I}} \tilde{q}^{\prime \dot{I}} \right)}\;,\nonumber\\
  V &= \int_\Sigma\!\!d\sigma \!\! \int \!\frac{du}{u^3} \;\;\, w(u) \;\; \bar{\delta}^{2}\!\left(u \lambda_{\alpha}-\epsilon_{\alpha}\,\right) \; e^{u \left(\mu^{\dot\alpha} \tilde{\epsilon}_{\dot\alpha} + \eta^{\prime \dot{I}} q_{\dot{I}} + \eta^{I} q^\prime_I \right)}\;,\\
  \tilde{V} &=  \int_\Sigma\!\!d\sigma \!\! \int \!\frac{du}{u^3} \;\;\, \tilde{w}(u) \;\; \bar{\delta}^{2}(u \tilde\lambda_{\dot\alpha}-\tilde\epsilon_{\dot\alpha})\;\;  e^{u \left(\tilde{\mu}^{\alpha} \epsilon_{\alpha} + \tilde{\eta}^\prime_I \tilde{q}^I + \tilde\eta_{\dot{I}} \tilde{q}^{\prime \dot{I}} \right)}\;,\nonumber
  \end{align}
  where for fixed vertex operators $\mathcal{W}(u)$ and $\tilde{\mathcal{W}}(u)$ are again products of fermionic ghost fields and delta functions of bosonic ghost fields, and for integrated vertex operators  
  \begin{align}
  \label{vo_2_massless}
    w(u) =& \; \left(u[\tilde{\lambda} \tilde{\epsilon}] - u^2 \,[\tilde{\epsilon} \rho_2]\,[\tilde\epsilon \tilde{\rho}_2]\right) \left(u\,\tilde{\eta}_{\dot{I}} \Omega^{\dot{I} \dot{J}} q_{\dot{J}} - u^2 \,\, q_{\dot{I}} \tau_{2}^{\dot{I}}\,\,q_{\dot{J}} \Omega^{\dot{J} \dot{K}} \tilde{\tau}_{2 \dot{K}}  \right)_{q_{\dot{I}} \ne 0}\nonumber\\
   \tilde{w}(u) =& \;\left(u \!\braket{\lambda \epsilon} \!- u^2 \!\!\braket{\epsilon \rho_1}\!\braket{\epsilon \tilde{\rho}_1}\right) \left(u\,\eta^I \Omega_{IJ} \tilde{q}^J\, - u^2 \,\,\tilde{q}^I \!\tau_{1I}\,\, \tilde{q}^J \Omega_{JK} \tilde{\tau}_1^J \, \right)_{\tilde{q}^I \ne 0}\,.
  \end{align}
  In contrast to the model in the previous section, $q^\prime$ and $q$ are distinct parameters and also $\tilde{q}^\prime$ and $\tilde{q}$ because of the SU(4) $\!\otimes\!$ SU(4) R-symmetry. Further, by setting $q_I \!\!=\!\! 0$ and $\tilde{q}^I \!\!\!=\!\! 0$ and omitting the new factors in \eqref{vo_2_massless} involving the $\eta$ and $\tau$ fields one obtains simpler vertex operators including the ones representing gravitons.\\
    
        To get the tree scattering matrix, let $g$ denote the set of positive-helicity vertex operators $V^g$ that do not include the factor with $\tau_2$ fields in $w(u)$ in \eqref{vo_2_massless}, $\tilde{g}$ the set of negative-helicity vertex operators $\tilde{V}^g$ that do not include the factor with $\tau_1$ fields in $\tilde{w}(u)$ in \eqref{vo_2_massless}, $h$ the set of the other positive-helicity vertex operators $V$, and $\tilde{h}$ the set of other negative-helicity vertex operators $\tilde{V}$. Note that $g$ and $\tilde{g}$ include the vertex operators for the gravitons. Then a tree scattering amplitude looks like:
  \begin{equation*}
  \mathcal{A} = \left< \frac{1}{ \mathrm{vol} \, \mathrm{GL}(2, \mathbb{C})} \, \prod_{j \in \tilde{h}} \! \tilde{V}_j \prod_{p \in h} \! V_{p} \,\, \prod_{i \in \tilde{g}} \! \tilde{V}_i^g \prod_{k \in g} \! V_k^g \right>\,, 
  \label{amplitude-1}
  \end{equation*}
  where the factor $\mathrm{vol} \, \mathrm{GL}(2, \mathbb{C})$ comes from three zero-modes of the $c$ ghost and one degree of freedom remaining from gauging the little group \cite{Albonico:2020}, and where one member of $h$ (or of $g$ when $h$ is empty) and one member of $\tilde{h}$ (or of $\tilde{g}$ when $\tilde{h}$ is empty) are fixed vertex operators\footnote{$g \cup h$ and $\tilde{g} \cup \tilde{h}$ are either both empty or non-empty \cite{Bianchi:2008pu, Grisaru:1977px, Grisaru:1976vm}.}, fixed for every worldsheet supersymmetry represented in the correlation function.\\
    
  The computation of the amplitude is standard \cite{Geyer:2014, Geyer:2018, Albonico:2020, Geyer:2020, Albonico:2022, Albonico:2023} except for the product of factors containing the auxiliary $\tau$ fields leading to reduced determinants of fermionic Hodges-like matrices, similarly to the bosonic fields. Explicitly:\\
  \begin{equation}
  \mathcal{A} = \int \!\!\!  \prod_{\genfrac{}{}{0pt}{}{i \in g \cup h }{ \cup \tilde{g} \cup \tilde{h}}} \!\!\! \frac{\ud u_i} {u_i^3} \!\!
  \prod_{r \in g \cup h} \!\!\! \ud \sigma_r \,\bar\delta^2\!(\epsilon_r \! - \! u_r \lambda(\sigma_r)) \!\!\! \prod_{l \in \tilde{g} \cup \tilde{h}} \!\!\!  \ud \sigma_l \,\bar\delta^2\!(\tilde{\epsilon}_l \! - \! u_l \tilde{\lambda}(\sigma_l))
  \frac{\mathrm{det}^{\prime} H \,\, \mathrm{det}^{\prime} \tilde{H} e^{F_{\mathcal N}}}{ \mathrm{vol} \, \mathrm{GL}(2, \mathbb{C})}  \mathrm{G}^{\prime}
  \label{tree-amplitude}
  \end{equation}
    where $\lambda(\sigma_r)$ and $\tilde{\lambda}(\sigma_l)$ fulfill the scattering equations
  \begin{equation*}
  \begin{aligned}
  &\lambda(\sigma_r) = \sum_{l \in \tilde{g} \cup \tilde{h}} \frac{u_l \epsilon_l}{\sigma_r - \sigma_l}, &\tilde{\lambda}(\sigma_l) = \sum_{r \in g \cup h} \frac{u_r \tilde{\epsilon}_r}{\sigma_l  - \sigma_r},
  \end{aligned}
  \end{equation*}
  $H$ is a symmetric $(|g| \!+\! |h|) \times (|g| \!+\! |h|)$ matrix, and $\tilde{H}$ is a symmetric $(|\tilde{g}| \!+\! |\tilde{h}|) \times (|\tilde{g}| \!+\! |\tilde{h}|)$ matrix
  arising from the contractions between the twistor and $\rho$ fields appearing in the
  vertex operators with the following elements:
  \begin{equation*}
  \begin{aligned}
  &H^{lr} = \frac{u_l u_r [\tilde{\epsilon}_l \tilde{\epsilon}_r]}{\sigma_l - \sigma_r} \;\, \text{for} \; l,r \!\in\! g \cup h, l \!\ne\! r, &H^{ll} \, = - \!\!\!\!\! \sum_{r \in g \cup h \setminus \{l\}} \!\! H^{lr} \; \text{for} \; l \!\in\! g \cup h\,,\\
  &\tilde{H}^{lr} = \frac{u_l u_r \braket{\epsilon_l \epsilon_r}}{\sigma_l - \sigma_r} \; \text{for} \; l,r \!\in\! \tilde{g} \cup \tilde{h}, l \!\ne\! r, &\tilde{H}^{ll} \, = - \!\!\!\!\! \sum_{r \in \tilde{g} \cup \tilde{h} \setminus \{l\}} \!\! \tilde{H}^{lr}  \; \text{for} \; l \!\in\! \tilde{g} \cup \tilde{h}\,.\\
  \end{aligned}
  \label{Matrices}
  \end{equation*}
  $H$ and $\tilde{H}$ each have co-rank one with vanishing determinant and det$^{\prime}$ indicates the operation of removing one row and one column before computing the determinant. The result of this operation is actually independent of the choice of row and column removed.\\

 The supersymmetric exponential factor exp $F_\mathcal{N}$ becomes like \eqref{super-factor} in the previous section, except for increased super polarization data $q^{\mathcal{I}}$ with $\mathcal{N} = 8$ components:
  \begin{equation*}
  \label{super-factor-8}
  e^{F_\mathcal{N}} = \mathrm{exp} \!\left[ \;\sum_{\genfrac{}{}{0pt}{}{j \in \tilde{g} \cup \tilde{h}}{k \in g \cup h}} \frac{u_j u_k}{\sigma_j \!-\! \sigma_k} \tilde{q}^{\mathcal{I}}_j q_{k \mathcal{I}} \right]\,, \,q_{\mathcal{I}} = (q^{\prime}_I, q_{\dot{I}}), \tilde{q}^{\mathcal{I}} = (\tilde{q}^I, \tilde{q}^{\prime \dot{I}}), \, q_{\mathcal{I}} \!\equiv\! \Omega_{\mathcal{IJ}} \tilde{q}^{\mathcal{J}}\,.
  \end{equation*} 
  This is the conventional massless limit of the supersymmetric exponential factor \cite{Albonico:2020}.\\
  
  G$^\prime$ is 1 for empty $h$ and $\tilde{h}$, otherwise stands for the product of reduced determinants $\mathrm{det}^\prime G_{\tilde{h}} \, \mathrm{det}^\prime G_{h}$ of the fermionic Hodges-like matrices $G_{\tilde{h}}, G_h$ defined by
  \begin{equation}
  \begin{aligned}
  &G_h^{lr} = \frac{u_l u_r}{\sigma_l - \sigma_r} q_{l \dot{I}} \Omega^{\dot{I} \dot{J}} \!q_{r \!\dot{J}}\,, \, l \!\ne\! r \in h, &\;G_{h}^{ll} \, =\! -\!\! \sum_{r \ne l \in h} \! \frac{u_l u_r }{\sigma_l - \sigma_r} q_{l \dot{I}} \Omega^{\dot{I} \dot{J}} \!q_{r \!\dot{J}}\,,\\ 
  &G_{\tilde{h}}^{lr} = \frac{u_l u_r}{\sigma_l - \sigma_r} \tilde{q}_l^I \Omega_{IJ} \tilde{q}_r^J\,, \;\; l \!\ne\! r \in \tilde{h}, &\;G_{\tilde{h}}^{ll} \, =\! -\!\! \sum_{r \ne l \in \tilde{h}} \! \frac{u_l u_r }{\sigma_l - \sigma_r}\; \tilde{q}_l^I \Omega_{IJ} \tilde{q}_r^J\,.
  \end{aligned}
  \label{G-Matrix-1}
  \end{equation} 
  
  $G_h$ and $G_{\tilde{h}}$ have co-rank 1 and the reduced determinant det$^\prime G_m$ is basically the product of all but one diagonal element of $G_m$ with all terms cancelled that include some propagator loop of elements in the $m$ set. This means, for instance, that terms in the scattering amplitude of spin 1 particles with propagator loops containing only helicities of the same sign are forbidden. This is known to be valid for gluon scattering in QCD \cite{Adamo:2015}. When $g \cup \tilde{g}$ only contains gravitons and all particles in $h \cup \tilde{h}$ have spin $\le 1$ then a single trace term will be a propagator loop with two missing links typically between each of the two fixed vertex operators and some other location which are filled by pulling down the appropriate propagators from the exponential $e^{F_{\mathcal{N}}}$. Such a single trace loop constitutes a kinematic Parke-Taylor (PT) factor:
  \begin{equation*}
  \label{Parke-Taylor}
  \mathrm{PT} = N / \left((\sigma_{|h| + |\tilde{h}|} - \sigma_1)\!\!\! \prod_{r < s \in h} \!\!\!(\sigma_r \!-\! \sigma_s) \;\; (\sigma_{|h|} - \sigma_{|h| + 1}) \!\!\!\prod_{|h|+r < |h|+s \in \tilde{h}} \!\!\!(\sigma_{r + |h|} \!-\! \sigma_{s + |h|}) \right)\,,
  \end{equation*}
  where $N$ stands for a numerator consisting of a single trace product of Grassmann odd parameters $q_{\mathcal{I}}$ and where the products run through an ordered union of $h$ and $\tilde{h}$, respectively. For spin $\frac{1}{2}$ and scalar particles higher powers in $q$ components need to be considered.\\
  
  Although this model looks promising, it also has some deficiencies. The treatment of the $\eta^{\mathcal{I}}$ and $\tilde{\eta}_{\mathcal{I}}$ components is not fully symmetric and the generalization to the massive case is unknown\footnote{In order to make progress in this direction, one might have to switch the representation to use fermionic delta functions and scattering equations for the fermionic fields \cite{Albonico:2020}, but then lose the convenience of a supersymmetric exponential factor in the scattering amplitudes.}. Like in the previous section the order of the PT factor is not arbitrary: the same sign helicities need to be adjacent to each other. This could be overcome by exploiting the factorization properties of the amplitude. As shown in \cite{Adamo:2015}, the amplitude \eqref{tree-amplitude}, when $G^\prime$ is a PT factor, factorizes into smaller tree-amplitudes. In particular, the PT factor factorizes by itself and, therefore, it is possible to make the helicity order in the PT factor arbitrary by gluing together smaller $n_L$ and $n_R$ point amplitudes to a $n = n_L + n_R - 2$ point tree amplitude accordingly. Admittedly, this procedure to circumvent the obstruction looks somewhat contrived.\\
        
  In conclusion, for pure graviton scattering the model in this section gives the expected graviton tree scattering amplitude \cite{Geyer:2014, Albonico:2023} and the main difference with a SYM tree scattering amplitude for spin $\!\le\! 1$ particles consists of the additional reduced determinants det$^{\prime} H\!\!$ det$^{\prime} \tilde{H}$ indicating that all particles exchange gravitons with each other, even without any external gravitons present. For spin $\le 1$ particles there might be even more differences due to the general form of the fermionic Hodges-like matrices \eqref{G-Matrix-1}.\\
  
  The actual calculation of an n-point amplitude can proceed by solving the scattering equations and inserting a solution into the various parts of the integrand including the Jacobian. This gives a power series in the Grassmann odd parameters $q_\mathcal{I}$ reflecting the $\mathcal{N} = 8$ supersymmetry, and the coefficients describe scattering of particles of different spin.\\

\section{Summary and Outlook}
  \label{Summary}  
  This work examined the spectrum of the massive ambitwistor supergravity model found in \cite{Geyer:2020, Albonico:2022, Albonico:2023} based on oscillator expansions. It turned out that the spectrum is covered by Table \ref{RSpectrumTable1} reflecting $\mathcal{N} = 8$ supersymmetry. As the twistors are worldsheet spinors there is a R sector and a NS sector to consider and both have the same particle spectrum except in the NS sector it is truncated because of the constraint that physical states have at most two on-shell oscillator modes.\\
    
  The massive models considered in \cite{Geyer:2020, Albonico:2022, Albonico:2023} have a vanishing little group anomaly coefficient and also a zero Virasoro central charge after a contribution from compactifying extra dimensions. In this article, the supergravity model got first extended by doubling and gauging the fermionic twistor components with help of auxiliary fields in such a way that the resulting model was anomaly-free and the scattering amplitudes for spin $\le 1$ particles exhibited a kinematic (no color trace) PT factor like a SYM model but without an external current algebra. The spectrum of the first modification which was limited to massless particles showed similarity to the spectrum of the Berkovits-Witten model though without the disturbing 'dipole' modes. Also, it is the same spectrum as of another model found by the author earlier, described in appendix \ref{Appendix}. Unfortunately, for both models the scattering amplitudes can look complicated and difficult to handle. Then, a much improved, still massless model variation was presented with the same spectrum as the original massive model and scattering amplitudes that differ from the massless limit of the original model only by a factor consisting of two reduced determinants of fermionic Hodges-like matrices with co-rank 1 that under special circumstances can lead to a Parke-Taylor factor.\\
  
  The issue about the truncated spectrum in the NS sector can be argued away. This is based on the observation that a little group SL(2$,\mathbb{C}$) gauge transformation can change the pair of supertwistors from the R sector to the NS sector and vice versa, for instance by multiplying one supertwistor with $\sigma^{\frac{1}{2}}$ and the other one with $\sigma^{-\frac{1}{2}}$. Thus, from a classical point of view, the NS and R sector are equivalent in the two-twistor models. Therefore, it is sufficient to quantize only the more convenient R sector. On the quantized level, the NS sector can then be regarded as resulting from a spectral flow operation on the R sector, with the $\pm\frac{1}{2}$ modes in the NS sector becoming generalized zero modes\footnote{For spectral flow of twistor modes in the context of ADS/CFT duality, see for instance \cite{Gaberdiel:2021jrv}.}, overcoming the truncation constraint of the NS spectrum. To have equivalent spectrum before and after spectral flow and to be able to identify the spectrum in the NS sector with the one in the R sector, this interpretation works exactly when for the $L_0$ constant $a = 0$ in the R sector and $a = 1$ in the NS sector, i.e. for the original massive model and the model in section \ref{Improved} but not for the intermediate model in section \ref{Intermediate} or the model in appendix \ref{Appendix}.\\
  
  Concentrating on the spectrum of $\mathcal{N} = 8$ supersymmetry as shown in Table \ref{RSpectrumTable1} and comparing modes in the triangles above and below diagonal lines drawn from upper left to right bottom, then the fermionic (half-integer spin) modes in the lower triangle represent antiparticles of the ones in the upper triangle in both halves of the table. Further, the 6 representation of SU(4) can be conveniently decomposed into a SU(2) $\!\otimes\!$ SU(2) $\!\otimes\!$ U(1) basis \cite{Piepenbring:1987} where out of a sextuple the first two elements and the next two can be used each as a duplet for a fundamental representation of SU(2) leaving the last two elements as a third duplet although not representing an SU(2). Nevertheless, this allows to consider $(\pm\frac{1}{2}|4 \!\otimes\! 6)$ and $(\pm\frac{1}{2}|\bar{4} \!\otimes\! 6)$ as 3 generations of the spin $\frac{1}{2}$ content of the Pati-Salam model \cite{Pati:1974}. Also, it is intriguing to see that the spectrum contains $(\pm1|4 \!\otimes\! \bar{4}) = (\pm1|15 \!\oplus\! 1)$ exactly once that is available for the adjoint vector bosons of SU(4) and also two $(\pm1|1 \otimes 6)$ that have each a decomposition that includes an adjoint representation of SU(2)\cite{Feger:2019tvk} \footnote{The fact that the N=8 supergravity spectrum contains exactly the spin $\frac{1}{2}$ content for all the fermions of the standard model and also for 8 massive gravitinos is well known \cite{Meissner:2014joa, Meissner:2018gtx}.}.\\  
  
For future work, when disregarding graviton exchange, i.e. setting the reduced determinants det$^{\prime} H\!\!$ det$^{\prime} \tilde{H}$ equal to 1, genus 0 scattering amplitudes of the model in section \ref{Improved} should be checked for how far they match gluon and qcd tree amplitudes from the literature. For instance, what is immediately apparent, single trace N$^k$MHV amplitudes look the same as in SYM, apart from color traces.\\
 
 Other open questions for the improved model in section \ref{Improved} are a massive generalization, loop scattering amplitudes for genus $\!\ge\!1$ worldsheets, modular invariance, and unitarity. The ultimate goal, of course, would be to show that it leads to a consistent UV-complete string field theory in twistor space with a classical limit that can be compared with general relativity.\\
 
 The above solution for the truncated spectrum in the NS sector deserves more examination, working out more details about the spectral flow operation.\\

 Further, the physical meaning of the auxiliary fields seems to be a mystery. They got introduced in an ad hoc fashion, as a kind of supertwistors with reversed statistics to support worldsheet supersymmetries. Without showing up in physical states they are on the level of ghosts but if they are just a matter of mathematical convenience what are the deeper lying principles behind their origin? Maybe they are superfield components in a superspace although the superfield would look rather complicated because of the many supersymmetries and other worldsheet symmetries breaking the supertwistors apart. On the other hand, if they are remnants of compactifying extra dimensions the exact reduction mechanism would need to be investigated.\\

  \appendix
  \section{Non-Supersymmetric Model} 
  \label{Appendix}
  
  Here the intermediate model of section \ref{Intermediate} gets changed to one that gauges worldsheet supersymmetries without the use of (mysterious) auxiliary fields.
  The model considered here is, in the notation of section \ref{Intermediate},
   \begin{equation*}
  \label{appendix-sugra-action}
S =\int_\Sigma \mathcal{Z}^a \cdot \bar\partial \mathcal{Z}_a + A_{ab} \, Z^a \cdot Z^b + S_g\,,
  \end{equation*}
  with
  \begin{equation*}
  \label{gauge-action}  
  S_g=\int_{\Sigma} F^{\alpha \dot\alpha}_{ab} \lambda_{\alpha}^a  \tilde{\lambda}_{\dot\alpha}^b + G_{ab}^{\alpha I} \lambda_{\alpha}^a \tilde{\eta}_I^b + \tilde{G}_{ab}^{\dot\alpha I} \tilde{\lambda}_{\dot\alpha}^a \eta_I^b \,,
  \end{equation*} 
  where the fermionic twistor components $\eta^a_I$ do not participate in the little group symmetry of the two twistors which breaks the supertwistors apart. $F^{\alpha \dot\alpha}_{ab}$ gauges the conformal translations and $G_{ab}^{\alpha I} , \tilde{G}_{ab}^{\dot\alpha I}$ gauge the superconformal translations.\\
  
  BRST quantization leads to $(b,c)$ ghosts for worldsheet gravity and:\\
  from the bosonic fields $\{A_{ab}, F^{\alpha \dot\alpha}_{ab}\}$ to fermionic ghosts $\{(M_{ab}, N_{ab}), (f^{\alpha \dot\alpha}_{ab}, e^{\alpha \dot\alpha}_{ab})\}$ and\\
  from the fermionic fields $\{G_{ab}^{\alpha I} , \tilde{G}_{ab}^{\dot\alpha I}\}$ to bosonic ghosts $\{(\beta_{ab}^{\alpha I}, \gamma_{ab}^{\alpha I}),(\tilde{\beta}_{ab}^{\dot\alpha I} \tilde{\gamma}_{ab}^{\dot\alpha I}\}$.\\
  
  The SL(2,$\mathbb{C}$) anomaly coefficient is:
  \begin{equation*}
  a_{sl2} = \frac{3}{2}(4_{\mathcal{Z}}) - 6 (4_{fe}) + \frac{3}{2}(2\mathcal{N}_{\beta \gamma} + 2\mathcal{N}_{\tilde{\beta} \tilde{\gamma}}) - 6_{M\! N} = -6(4 - \mathcal{N})\,.
  \end{equation*}
  The central charge is 
  \begin{equation*}
  c = (-8 + 2 \mathcal{N})_{\mathcal{Z}} - 26_{bc} - 6_{M\! N} - 32_{fe} + 8\mathcal{N}_{\beta \gamma} + 8\mathcal{N}_{\tilde{\beta} \tilde{\gamma}} = -18(4 - \mathcal{N})\,.
  \end{equation*}
  Therefore, the theory is anomaly-free for $\mathcal{N} = 4$. Then the $L_0$ constant $a$ is in both the NS and R sector given by
  \begin{equation*}
  24 a = (0)_{\mathcal{Z}} - 2_{bc} - 6_{M\! N} - 32_{fe} + 8\mathcal{N}_{\beta \gamma} + 8\mathcal{N}_{\tilde{\beta} \tilde{\gamma}} = 24\,,
  \end{equation*}
   i.e. $a = 1$, exactly like for the intermediate model in section \ref{Intermediate}. Although $\mathcal{N}$ has here half the value, it has the same R-symmetry because the index $a$ of $\eta^a_I$ is not part of the little group but of the R-symmetry. This means the two models have the same spectrum although the internal little group representation does not need to be the same across the R-symmetry multiplet when disregarding target space supersymmetry\footnote{This allows an interpretation of the spectrum to include 3 generations of the Pati-Salam model \cite{Kunz_1:2020}, similar to the model in section \ref{Improved}.}. The current model was originally found by the author using other notation \cite{Kunz:2020}, and explored more in \cite{Kunz_1:2020} providing vertex operators and scattering amplitudes. Because it suffers from similar issues as the model in section \ref{Intermediate}, it can be abandoned in favor of the improved model in \ref{Improved} that have more potential to be realistic.\\

   \appendix  

\bibliography{TwistorString}
\end{document}